\documentclass[prb,aps,eqsecnum]{revtex4}

\usepackage{amssymb}
\usepackage{amsmath}
\usepackage{tabularx}
\usepackage{pifont}
\usepackage{makeidx}
\usepackage[dvips]{graphicx}
\begin{document}
\bibliographystyle{apsrev}

\title{Electric field gradients from first-principles and point-ion
calculations}
\author{E. P. Stoll and P. F. Meier}

\affiliation{Physics Institute, University of Zurich, CH-8057 Zurich, 
         Switzerland}
\author{T. A. Claxton}
\affiliation{Department of Chemistry, University of York, York, YO105DD, UK}

\begin{abstract}
Point-ion models have been extensively used to determine ``hole numbers''
at copper and oxygen sites in high-temperature superconducting cuprate
compounds from measured nuclear quadrupole frequencies.  The present study
assesses the reliability of point-ion models to predict electric field 
gradients accurately and also the implicit assumption that the values can be
calculated from the ``holes'' and not the total electronic structure.
First-principles cluster calculations using basis sets centred on 
the nuclei have enabled the determination of the charge and spin density
distribution in the CuO$_2$-plane. The contributions to the electric field
gradients and the magnetic hyperfine couplings are analysed in detail. 
In particular they are partitioned into regions in an attempt to find
a correlation with the most commonly used point-ion model, the Sternheimer
equation which depends on the two parameters $R$ and $\gamma$.
Our most optimistic objective was to find expressions for these parameters,
which would improve our understanding of them, but although estimates of the
$R$ parameter were encouraging the method used to obtain the $\gamma$ parameter
indicate that the two parameters may not be independent.  
The problem seems to stem from the covalently bonded nature of the 
CuO$_2$-planes in these structures which severely questions using 
the Sternheimer equation for such crystals, since its derivation is 
heavily reliant on the application of perturbation theory to 
predominantly ionic structures. Furthermore it is shown that the 
complementary contributions of electrons and holes in an isolated ion 
cannot be applied to estimates of electric field gradients at copper 
and oxygen nuclei in cuprates. 
\end{abstract}

\pacs{PACS numbers: 71.15.Mb, 78.20.Bh, 61.72.Ji}
\maketitle


\section{Introduction}
There is a large quantity of nuclear magnetic resonance (NMR) data from
high-temperature superconducting cuprate crystals from which
electric field gradients (EFG) can be derived.  EFG's are a measure of the
non-spherical components of the charge distribution surrounding the nucleus of
interest and is used to estimate the hole population in models of
superconductivity. Most 
estimations~\cite{adrian,Shimizu,pennington,Garcia,kat,takigawa,hanzawa} have 
been made using a point-ion model with Sternheimer
correction factors, called here the Sternheimer equation~\cite{Stern,Stern2}
(SE) which is briefly discussed in Sec.~\ref{st}. 
In particular the measured changes of the EFG on doping
have been discussed~\cite{ohta,zheng,asayama,Kupcic,walstedt} 
in terms of the distribution of
the additional holes among the orbitals on each ion.
In related areas of 
research, however, point-charge models are apparently no longer
in use \cite{pc} although there does not seem to be any report in 
the literature discussing the 
unreliability of the SE.  In this paper we address this problem,
particularly for systems where there is evidence that the bonding between
``ions'' is more covalent than ionic.  The CuO$_2$ sheets, which are a common
feature of all cuprates showing high-temperature superconductivity
behaviour, are thought to have bonds showing a distinct covalent character.

In addition to the point-ion approximation, the above mentioned 
semi-empirical analyses of EFG values in cuprate superconductors are 
also based on the assumption that the EFG values
at a nucleus can be calculated (with opposite sign) for that configuration
where the unoccupied spin orbitals are assumed to be occupied
and the occupied spin orbitals are assumed to be unoccupied.
For the Cu $3d^9$ ion, in particular,  
it is expected that the EFG is just the same as for $3d^1$ but with opposite
sign. It seems that this concept is
widely adopted unconditionally. In the present paper we also address
this assumption, called here the electron-hole symmetry. We demonstrate
that it is entirely unjustified for the Cu ions in a cuprate environment
but also leads to false
estimates of the EFG values at the oxygens in the CuO$_2$-planes.

In order to quantify our doubts on the applicability of the SE 
and of the electron-hole symmetry to the evaluation of EFG in copper oxides
we have used the wave functions from previously published first-principles
calculations and complemented them for illustrative purposes with
additional simulations. These are cluster calculations~\cite{La,Y} 
on La$_2$CuO$_4$ and YBa$_2$Cu$_3$O$_7$ cuprates using the density functional 
theory with local density approximation and generalized gradient corrections,
which have provided EFG data for the Cu and O in the CuO$_2$-planes
in agreement with experiment. Calculations which used augmented plane waves 
give similar agreement~\cite{pw1,pw2} for all nuclei except copper in the
CuO$_2$ plane. It is not the purpose of the present paper to discuss such 
differences since the focus is solely on the reliability of point-ion 
calculations. For detailed comparisons with experimental measurements we 
refer to Refs~\cite{La,Y}.

The general idea of the cluster approach to electronic structure calculations 
of properties which depend upon predominantly local electron densities
is that the parameters that characterize a small cluster should be 
transferable to the solid and largely determine its properties. 
The essential contributions to EFGs and to magnetic hyperfine fields are 
given by rather localized interactions and therefore it is expected that 
these local properties can be determined and understood with clusters 
calculations. Approximations must be made concerning the treatment of the
lattice in which the cluster is embedded. Using as large a cluster 
as is possible is of course advantageous. It is necessary, however, that 
the results obtained should be checked with respect to their dependence 
on the cluster size.

The basic principles of cluster calculations are briefly discussed 
in Sec.~\ref{cluster} and the general contributions to the one-electron 
operator from regional partitioning are given in Sec.~\ref{rp}. 
In Sec.~\ref{hfc}, this is then applied to the EFG operator
and also to the hyperfine coupling operator. The latter
which can be used for clusters with unpaired electron spins
is very similar to the EFG operator. 
The only difference is that it uses the spin density and the EFG the
charge density.

Effectively the cluster calculation of the EFG is divided into contributions
from the ion of interest (the target ion), the rest of the cluster and the
overlap between these two.
This regional partitioning technique
is described in Sec.~\ref{exrp} and the contributions to the
EFG for a particular cluster calculation are given as an example.
In Sec.~\ref{fpp} correlations of first-principles partitions with 
Sternheimer terms are investigated.
The above mentioned partition enable the Sternheimer antishielding factors, 
R and $\gamma$ 
to be associated with quantities calculated from first principles (see
Sec.~\ref{identification}), which in
turn allows us to compare the predictions of the SE for a model cluster with
that of the {\it ab initio} calculation of the same cluster. 
This provides
a much more sensitive test of the SE than could otherwise be obtained.
In Sec.~\ref{RegionI} the contributions from the
target ion are analysed in terms of the individual orbitals indicating how
the ``holes'' have been determined.  The shortcomings of the simplifying
approaches are pointed out. It is shown that the values of the EFG are
determined by a subtle cancellation of large individual terms.
In Sec.~\ref{ehs} and \ref{DMA} the electron-hole symmetry is studied. 
A summary and conclusions are given in Sec.~\ref{summary}.

Except for energies, atomic units are used throughout, i.e. the 
EFG components $V_{ii}$ are given in 
$e\,a_B^{-3} = -\!\!\mid\!e\!\!\mid\!\!a_B^{-3} = -$Ha/$a_B^2 $. The quantity
$q_{ii} = V_{ii} /\! \mid\!e\!\!\mid $ then corresponds 
to $-9.7174 \times 10^{21}$ in units of V/$m^2$.

\section{The Sternheimer Equation}
\label{st}
The Sternheimer equation has been written in the following form\cite{Stern}:
\begin{equation}
V_{ii}=(1-R)V_{ii}^{local}+(1-\gamma)V_{ii}^{lattice}
\label{vstern}
\end{equation}
where $V_{ii}$ is one of the diagonal components of the EFG tensor
for a target ion which
can be determined experimentally, $V_{ii}^{local}$ is the experimental EFG
component of the target free ion and $V_{ii}^{lattice}$ is the contribution
to the EFG component from the charges in the lattice surrounding the target
ion.  The two parameters in the equation are both antishielding factors; R
arises from the electrons in the valence shell of the target ion which are
possibly overlapping the electron distribution of the nearest neighbour ions,
and $\gamma$ accounts for the contribution of the EFG due to the polarisation
of the target ion in the electric field of the environmental charges.

The crystal structure is therefore split up into three regions, the isolated
target ion (here referred to as {\it local}), the space between the target ion and its nearest neighbours, the
rest of the ions in the crystal (here referred to as {\it lattice}).

There is large spread  of values for the parameters, in particular $\gamma$,
derived from EFG data for superconducting cuprates. For copper, values 
for $\gamma$ of -7.6, -10.4, -17, and -20 have been reported in 
Refs.~\cite{Garcia,pennington,adrian,Shimizu}, respectively.
For planar oxygen ions the lattice contributions 
$(1-\gamma)V^{lattice}_{xx}$ accounts for 36\% and 60\% of the total 
$V_{xx}$ in Ref.~\cite{Kupcic} and Ref.~\cite{walstedt}, respectively.

\section {First Principles Calculations on Clusters}

\subsection{Description of cluster calculations}
\label{cluster}
A cluster is a careful selection of ions within a crystal which are intended to be
able to calculate localised properties accurately.  The target ion (the ion whose
properties are to be calculated) should be at or very near the centre of the cluster.
The target and normally, at least, its nearest neighbours form the core of the cluster
and are treated most accurately using first-principles all-electron methods.
Outside this core the next shell of positively charged ions are represented by pseudopotential functions
which have been shown to behave better than just bare charges to represent the ions
since pseudopotentials prevent unrealistic electron density distortions
characteristic of the positive point charges.  Bare charges ($\approx$ 2000) are used outside
the shell of pseudopotentials to simulate the rest of the crystal lattice.
Some of the more remote charges from the target ion are moved slightly so that the target
ion experiences the correct Madelung potential.

The present work aims at an assessment of calculations
of EFG with a point-ion model and Sternheimer corrections and comparison
with first-principles methods. To illustrate the problems we use
here the results from three different clusters which all simulate the
compound La$_2$CuO$_4$ but we note that similar results have been obtained
for YBa$_2$Cu$_3$O$_7$. The core of these clusters comprise 
one, two and nine copper ions, respectively, each with an appropriate
number of nearest neighbour oxygen ions.

Only the central Cu ion in the cluster CuO$_6$/Cu$_4$La$_{10}$ 
(Fig.~\ref{clusterfig}(a) where X=Y=La) is used as the target ion which, 
together with the 6 nearest neighbour oxygen
ions, forms the core of the cluster used for the all-electron calculation.
The neighbouring 4 Cu and 10 La ions are represented by 
pseudo-potential functions. In the Cu$_2$O$_{11}$/Cu$_6$La$_{16}$ cluster 
(Fig.~\ref{clusterfig}(b) where X=Y=Z=La) we have used both the central
oxygen ion and the neighbouring two Cu ions separately as target ions.
The core of the cluster additionally includes the 10 nearest neighbour 
oxygen ions. The adjacent 6 Cu and 16 La ions are represented 
by pseudo-potential functions.

The nuclear positions have been chosen~\cite{lattice} according to the
tetragonal structure of La$_2$CuO$_4$ (space group $I4/mmm$)
with a = b = 3.77 {\AA} and c = 13.18 {\AA}, with a Cu-O(a) distance of 2.40 
{\AA} and with a Cu-La distance of 4.77 {\AA}.

The cluster core uses a 6-311G basis set as provided by Gaussian 98.
The density functional method was used to obtain the wave functions from which
the EFG at the target ions were calculated. This
procedure has been used consistently by us since the wave functions which
are produced give calculated properties which are in agreement with
experimental values.

It should be noted that the lattice region of the Sternheimer 
equation includes all bare charges, the pseudo-potential ions 
and all the ions of the cluster core except the target ion.

\subsection{Contributions to the one-electron operator from regional partitioning}
\label{rp}

Let us consider a system of $N$ nuclear centres.  The $K^{th}$ centre, at site
$\vec{R}_K$, is the origin for $n_K$ basis functions which are mutually 
orthogonal. The $k^{th}$ basis function on site $\vec{R}_K$ is denoted by
$B_{K,k}(\vec{r}-\vec{R}_K)$. The total number of basis 
functions (atomic orbitals) is
\begin{equation}
n_c=\sum_{K=1}^N n_K.
\end{equation}
the $c$ in $n_c$ identifies that our system here is the  $core$ of a cluster.
The molecular orbitals (MO), $\phi$, of the system are orthogonal linear
combinations of the atomic orbitals.  We allow for two sets of MO's, one set
to hold electrons of $\alpha$-spin projection, the other set  $\beta$-spin
projection.  The $m^{th}$ MO of $\alpha$-spin projection is
\begin{equation}
\phi_{m,\alpha}(\vec{r})=\sum_{K=1}^N \phi_{m,\alpha}^K (\vec{r}-\vec{R}_K) =
\sum_{K=1}^N \sum_{k=1}^{n_K}c^{K,k}_{m,\alpha} B_{K,k}(\vec{r}-\vec{R}_K) \label{MO}
\end{equation}
where the $c$'s are the MO coefficients.

The expectation value of any quantity, corresponding to the operator ${\cal{O}}(\vec{r})$,
associated with a nuclear site ${\vec{R}}_J$ for the MO $\phi_{m,\alpha}(\vec{r})$, is given by the matrix element
\begin{equation}
M^{\cal{O}}_{m,\alpha}(\vec{R}_J)=<\phi_{m,\alpha}(\vec{r})|{\cal O}(\vec{r}-\vec{R}_J)|\phi_{m,\alpha}(\vec{r})>.  \label{expect}
\end{equation}

Developing the MO's according to Eq. (\ref{MO}) we get
\begin{equation}
M^{\cal{O}}_{m,\alpha}(\vec{R}_J)=\sum_{K}\sum_{L}\Gamma^{\cal{O}}_{m,\alpha}(K,L)\label{tt}
\end{equation}
where, for convenience, we have defined:
\begin{equation}
\Gamma^{\cal{O}}_{m,\alpha}(K,L)=\sum_k^{n_K}\sum_l^{n_L}c_{m,\alpha}^{K,k}
c_{m,\alpha}^{L,l}<B_{K,k}(\vec{r}-\vec{R_K})|{\cal O}(\vec{r}-\vec{R}_J)|B_{L,l}(\vec{r}-\vec{R_L})> .
\end{equation}

We note that in Eq.~(\ref{tt}) $K$ and $L$ sum over all nuclear centres 
of the cluster and that the target ion is $J$.  The SE concentrates 
entirely on the target ion and so will we. Hence in Eq.~(\ref{tt}) we 
separate out the target ion as follows:
\begin{eqnarray}
M^{\cal{O}}_{m,\alpha}(\vec{R}_J)&=&\Gamma^{\cal{O}}_{m,\alpha}(J,J)
+\sum_{K\ne J}\Gamma^{\cal{O}}_{m,\alpha}(K,J)\nonumber\\
&&+\sum_{L\ne J}\Gamma^{\cal{O}}_{m,\alpha}(J,L)+\sum_{K\ne J}\sum_{L\ne
J}\Gamma^{\cal{O}}_{m,\alpha}(K,L)\label{Mexpectb}\\
&=&\,^{\rm I}\!M^{\cal O}_{m,\alpha}(\vec{R}_J)+\,^{\rm II}\!M^{\cal O}_{m,\alpha}(\vec{R}_J)
+\,^{\rm III}\!M^{\cal O}_{m,\alpha}(\vec{R}_J)\label{Mexpectc}
\end{eqnarray}

noting that $\,^{\rm II}\!M^{\cal O}_{m,\alpha}(\vec{R}_J) = \sum_{K\ne
J}\Gamma^{\cal O}_{m,\alpha}(K,J)
+ \sum_{L\ne J}\Gamma^{\cal O}_{m,\alpha}(J,L)$.

This results in the identification of three different terms.
\begin{enumerate}
\item  The first term comprises all contributions from on-site basis functions 
(that is, all basis functions centred at the target ion $\vec{R}_J)$ and 
is denoted by regional partition I.
\item The second and third terms in Eq.~(\ref{Mexpectb}) are numerically
identical and contain contributions
arising from {\it both} on-site and off-site ($\vec{R}_K,K\ne J$) basis 
functions (corresponding to regional partition II) and denoted by II in
Eq.~(\ref{Mexpectc}).
\item 
$^{\rm III}\!M^{{\cal O}}_{m,\alpha}(\vec{R}_J)$ contains {\it no} reference 
to the on-site basis functions. 
\end{enumerate}

This is illustrated schematically in 
Fig.~\ref{CudOp}.

If $\cal{O}$ is the identity operator {\it 1} then
\begin{eqnarray}
\sum_{K}\sum_{L}\Gamma^{\it 1}_{m,\alpha}(K,L)&=&\sum_{K,L}\sum_k^{n_K}\sum_l^{n_L}c_{m,\alpha}^{K,k}
c_{m,\alpha}^{L,l}<B_{K,k}(\vec{r}-\vec{R_K})|B_{L,l}(\vec{r}-\vec{R_L})>\nonumber\\
&=&\,^{\rm I}\!M^{{\it 1}}_{m,\alpha}(\vec{R}_J)+\,^{\rm II}\!M^{{\it 1}}_{m,\alpha}(\vec{R}_J)
+\,^{\rm III}\!M^{{\it 1}}_{m,\alpha}(\vec{R}_J)
\end{eqnarray}
\noindent
and since the basis orbitals (functions) on each centre have been 
conveniently chosen to be orthogonal
\begin{equation}
^{\rm I}\!M^{{\it 1}}_{m,\alpha}(\vec{R}_J)
=\sum_k^{n_J}(c_{m,\alpha}^{J,k})^2=\, ^{\rm I}\!N_{m,\alpha}(J).
\label{unit}
\end{equation}

If we define an overlap integral as
$S_{K,k,L,l}=<B_{K,k}(\vec{r}-\vec{R_K})|B_{L,l}(\vec{r}-\vec{R_L})>$
we get
\begin{equation}
^{\rm II}\!M^{{\it 1}}_{m,\alpha}(\vec{R}_J)=
2\sum_{K\ne J}\sum_k^{n_K}c_{m,\alpha}^{J,k}c_{m,\alpha}^{K,l}S_{J,k,K,l}
= \, ^{\rm II}\!N_{m,\alpha}(J).
\end{equation}

Whereas $^{\rm I}\!N_{m,\alpha}(J)$ has been interpreted by Mulliken
as the charge
on atom $J$ due to $\alpha$-spin electrons in MO $\phi_{m,\alpha}$, 
$^{\rm II}\!N_{m,\alpha}(J)$ is the $\alpha$-spin density that atom $J$
shares with all its neighbours in the same MO. With the definition
\begin{equation}
\rho_\alpha (\vec{R}_J) = =\sum_m^{occ} \rho_{m,\alpha} (\vec{R}_J)
=\sum_m^{occ}\left[^{\rm I}\!M^{{\it 1}}_{m,\alpha}(\vec{R}_J)+\frac{1}{2}\,^{\rm II}\!M^{{\it 1}}_{m,\alpha}(\vec{R}_J)\right]\label{mulli}
\end{equation}
(noting that the sum is only over occupied MOs only) this gives the Mulliken
charge density attributed to the nuclear centre at $\vec{R}_J$:
\begin{equation}
\rho^{Mull}(\vec{R}_J)=\rho_\alpha (\vec{R}_J)+\rho_\beta (\vec{R}_J).
\end{equation}


It should be noted that the Mulliken analysis of the charge distribution
has no physical meaning but it is very useful when discussing the charge
distribution in molecules and clusters, perhaps more suited to systems 
which are non-ionic rather than ionic.

\subsection{The EFG and hyperfine coupling operators}
\label{hfc}

The concepts developed above are now applied to expectation values of the
operator
\begin{equation}
{\cal{D}}^{ij}(\vec{x}) = (\nabla_i \nabla_j - \frac{1}{3}\delta_{ij}
\Delta)\frac{1}{x} - \frac{2}{3}\delta_{ij}\Delta\frac{1}{x}. 
\label{dijnabla}
\end{equation}
The expectation values of this operator cover three contributions of interest:
\begin{enumerate}
\item The Fermi contact density.

This is just the second term on the right hand side of
Eq.~(\ref{dijnabla}) and its expectation value gives rise to the expression 
$\frac{8 \pi}{3} | \psi (0) 
|^2$, where $|\psi(0)|^2$ is the spin density at the target nucleus. 
We will not discuss this term in the following but we note that the same 
analysis which will be performed for the contributions to the EFG has also 
been applied to the contact density. The corresponding results are 
given in Appendix~\ref{contactdensities}.

\item EFG operator

This operator is only the first term in Eq.~(\ref{dijnabla}) 
and transforms as a spherical harmonic of order 2 and its expectation value 
for an $s$-like charge or spin distribution vanishes. It is written as
\begin{equation}
{\cal{D}}^{ij}(\vec{x}) = \frac{3 x_i x_j - \delta_{ij} x^2}{x^5} .
\label{dijx5}
\end{equation}
The expectation values of
\begin{equation}
{\cal{D}}^{ij}_J = {\cal{D}}^{ij}(\vec{x} - \vec{R}_J )
\label{dijrj}
\end{equation}
determine the EFG tensor {\it only} if the total {\it charge} density distribution is used. 

\item Dipolar Hyperfine Coupling operator

This has exactly the same form as the EFG operator, but is {\it only} used with the
total {\it spin} density distribution.

\end{enumerate}

It will
be convenient, and unlikely to cause confusion since the only other operator 
defined is the unit operator {\it 1}, to replace $\cal{O}$ in 
equations (\ref{expect} - \ref{Mexpectc}) with just $ij$ when we should write 
${\cal{D}}^{ij}_J$ (there is no need to repeat J)

\begin{eqnarray}
M^{ij}_{m,\alpha}(\vec{R}_J)&=&<\phi_{m,\alpha}(\vec{x})|
{\cal{D}}^{ij}(\vec{x} - \vec{R}_J )
|\phi_{m,\alpha}(\vec{x})>\equiv M^{{\cal{D}}^{ij}}_{m,\alpha}(\vec{R}_J)\nonumber \\
&=&\,^{\rm I}\!M^{ij}_{m,\alpha}(\vec{R}_J)+\,^{\rm II}\!M^{ij}_{m,\alpha}(\vec{R}_J)
+\,^{\rm III}\!M^{ij}_{m,\alpha}(\vec{R}_J).\label{Dexpect}
\end{eqnarray}

Since the operator (15) contains a factor roughly proportional to the reciprocal
of the cube of the distance from nuclear centre $J$, we would expect only
those terms which describe the electron density close to the centre $J$ would be
significant. Clearly one of these terms is $^{\rm I}\!M^{ij}_{m,\alpha}(\vec{R}_J)$.  
Explicitly
\begin{equation}
^{\rm I}\!M^{ij}_{m,\alpha}(\vec{R}_J)=
\sum_k^{n_J}\sum_l^{n_J}c_{m,\alpha}^{J,k}c_{m,\alpha}^{J,l}D^{ij}_{J,k,J,l}(J)
\label{onecentre}
\end{equation}
where $D^{ij}_{K,k,L,l}(J)=
<B_{K,k}(\vec{x}-\vec{R_K})|{\cal{D}}^{ij}_J|B_{L,l}(\vec{x}-\vec{R_L})>$.
By summing over all the occupied MO's we define the quantities
\begin{equation}
^{\rm I}G^{ij}_\alpha=\sum_m^{occ} \!\,^{\rm I}\!M^{ij}_{m,\alpha}(\vec{R}_J)
\label{alpha}
\qquad \qquad
^{\rm I}G^{ij}_\beta=\sum_m^{occ} \!\, ^{\rm I}\!M^{ij}_{m,\beta}(\vec{R}_J) .
\end{equation}
The contribution of the ``on-site" terms (I) to the EFG is then given by the sum
\begin{equation}
^{\rm I}V^{ij}=\,^{\rm I}G^{ij}_\alpha+\,^{\rm I}G^{ij}_\beta\label{allonecentre}
\end{equation}
and the difference
\begin{equation}
^{\rm I}T^{ij}=\,^{\rm I}G^{ij}_\alpha-\,^{\rm I}G^{ij}_\beta\label{TI}
\end{equation}
\noindent is the corresponding contribution to the dipolar 
hyperfine tensor $T$.

Analogous definitions determine the mixed on-site and off-site contributions 
(II) and those of purely off-site contributions (III). This leads to the
following representations for the total EFG tensor
\begin{equation}
V^{ij}=\,^{\rm I}V^{ij}+\,^{\rm II}V^{ij}+
\,^{\rm III}V^{ij}+W^{ij} 
\label{EFGzus}
\end{equation}
\noindent
and for the dipolar term
\begin{equation}
T^{ij}=\,^{\rm I}T^{ij}+\,^{\rm II}T^{ij}+
\,^{\rm III}T^{ij} .
\label{HFCzus}
\end{equation}

Note that the last term in Eq.~(\ref{EFGzus}) represents the EFG 
contribution of all nuclear point charges $Z_K$. It is given by
\begin{equation}
W^{ij}=\sum_{K \neq J}\frac{3 (\vec{R}_K-\vec{R}_J)_i (\vec{R}_K-\vec{R}_J)_j
-\delta_{ij}|\vec{R}_K-\vec{R}_J|^2}{|\vec{R}_K-\vec{R}_J|^5} Z_K .
\label{efgpch}
\end{equation}
Of course $W^{ij}$ has no place in the electron-nuclear hyperfine coupling
tensor.







\subsection{Density Matrix Formulation of Partitioning}
\label{dmf}
The previous two subsections can be more succinctly described using 
density matrix terminology. In fact in Sec.~\ref{DMA} it enables certain 
conclusions to be
reached which would be difficult to achieve otherwise.

Let ${\bf B}$ be the column matrix of the complete set of basis functions
$B_{K,k}(\vec{r}-\vec{R_k})$.  It will be made clear shortly why we want 
to order the basis functions such that those belonging to the target 
ion are placed at the beginning.  The MO's of $\alpha$-spin 
projection can be written as the column matrix

\begin{equation}
\Phi_\alpha = {\bf c}^\dagger_\alpha {\bf B}\label{dm1}
\end{equation}
where ${\bf c}_\alpha$ is the matrix of the MO coefficients, each column
corresponding to a particular MO.
Equation (\ref{MO}) corresponds to the m$^{th}$ row of $\Phi_\alpha$ which can
be written here as
$\phi_{m,\alpha} = {\bf c}^\dagger_{m,\alpha}{\bf B}$
where ${\bf c}_{m,\alpha}$ is the m$^{th}$ column of ${\bf c}_\alpha$.

In order to obtain the expectation values of the operator ${\cal O}(\vec{r})$
it is convenient to define the matrix
\begin{equation}
{\bf b}^{\cal{O}}=<{\bf B}|{\cal O}(\vec{r})|{\bf B}^\dagger>\label{dm2}
\end{equation}
so that
\begin{equation}
M^{\cal O}_{m,\alpha}(\vec{R}_j)=Tr({\bf c}^\dagger_{m,\alpha}{\bf b}^{\cal{O}}{\bf c}_{m,\alpha})
=Tr({\bf c}_{m,\alpha}{\bf c}^\dagger_{m,\alpha}{\bf b}^{\cal{O}})
=Tr({\bf P}_{m}{\bf b}^{\cal{O}})
\end{equation}
which is equivalent to Eq.~(\ref{expect}).  ${\bf P}_{m}$ is the density matrix
${\bf c}_{m,\alpha}{\bf c}^\dagger_{m,\alpha}$.  In accordance with 
previous practice we have similarly for electrons with
$\beta$-spin 
projection ${\bf Q}_{m}={\bf c}_{m,\beta}{\bf c}^\dagger_{m,\beta}$.

Instead of discussing just individual MO's we can usefully define 
a {\it total} density matrix
for all the electrons with $\alpha$-spin projection as
\begin{equation}
{\bf P}= \sum_m^{occ}{\bf P}_{m}
={\bf c}_\alpha{\bf I}_\alpha{\bf c}^\dagger_\alpha
\end{equation}
where ${\bf I}_\alpha$ is a diagonal matrix with 1 on the m$^{th}$ diagonal
if MO $m$ is occupied and zero otherwise.

Since we have ordered the $n_c$ basis orbitals of the cluster core
such that all the $n_t$  basis orbitals on the $target$ ion
are listed first we partition the density matrix as follows,
\begin{equation}
{\bf P}_{m}= \left ( \begin{array}{ccc}
^{\rm I}{\bf p}_{m}&\vdots&^{\rm II}{\bf p}_{m}\\
\ldots&\ldots&\ldots\\
^{\rm II}{\bf p}_{m}^\dagger&\vdots&^{\rm III}{\bf p}_{m}\\
\end{array} \right )
\label{g1}
\end{equation}
where $^{\rm I}{\bf p}_{m}$
is an $n_t\times n_t$ matrix, $^{\rm II}{\bf p}_{m}$ is an $n_t\times (n_c-n_t)$ matrix
and $^{\rm III}{\bf p}_{m}$ is an $(n_c-n_t)\times (n_c-n_t)$ matrix.
If we define 
\begin{equation}
^{\rm I}{{\bf P}_{m}}= \left ( \begin{array}{ccc}
^{\rm I}{\bf p}_{m}&\vdots&{\bf _{n_t}0_{n_c-n_t}}\\
\ldots&\ldots&\ldots\\
\bf{_{{n_c}-{n_t}}0_{n_t}}&\vdots&\bf{_{n_c-n_t}0_{n_c-n_t}}\\
\end{array} \right ),
\label{g2}
\end{equation}
\begin{equation}
^{\rm II}{\bf P}_{m}= \left ( \begin{array}{ccc}
\bf _{n_t}0_{n_t}&\vdots&^{\rm II}{\bf p}_{m}\\
\ldots&\ldots&\ldots\\
^{\rm II}{\bf p}_{m}^\dagger&\vdots&\bf{_{n_c-n_t}0_{n_c-n_t}}\\
\end{array} \right )
\label{g3}
\end{equation}
and
\begin{equation}
^{\rm III}{\bf P}_{m}= \left ( \begin{array}{ccc}
\bf _{n_t}{\bf 0}_{n_t}&\vdots&\bf{_{n_t}0_{{n_c}-{n_t}}}\\
\ldots&\ldots&\ldots\\
\bf{_{{n_c}-{n_t}}0_{n_t}}&\vdots&^{\rm III}{\bf p}_{m}\\
\end{array} \right )
\label{g4}
\end{equation}
where ${\bf _j0_k}$ is a null $j\times k$ matrix, we have
\begin{equation}
{\bf P}_{m}=^{\rm I}{\bf P}_{m}+^{\rm II}{\bf P}_{m}
+^{\rm III}{\bf P}_{m} .
\end{equation}
Hence we have an equivalent expression for Eq.~(\ref{Mexpectc})
\begin{equation}
M^{\cal O}_{m,\alpha}(\vec{R}_J)=Tr(^{\rm I}{\bf P}_{m}{\bf b}^{\cal{O}})
+Tr(^{\rm II}{\bf P}_{m}{\bf b}^{\cal{O}})
+Tr(^{\rm III}{\bf P}_{m}{\bf b}^{\cal{O}})
\end{equation}
that is,  $^{\rm R}M^{\cal O}_{m,\alpha}(\vec{R}_J)
=Tr(^{\rm R}{\bf P}_{m}{\bf b}^{\cal{O}})$.

Having demonstrated the equivalence between the two mathematical approaches,
for example, Eq.~(\ref{alpha}) can be rewritten equivalently as
\begin{equation}
^{\rm I}G_\alpha^{ij}=Tr(^{\rm I}{\bf P}{\bf b}^{ij})~~~~~^{\rm I}G_\beta^{ij}
=Tr(^{\rm I}{\bf Q}{\bf b}^{ij}) .
\end{equation}
Thus the density matrix approach emphasises the partition method 
chosen diagrammatically.

Eqns.~({\ref{allonecentre}) and~(\ref{TI}) can be succinctly 
written as
\begin{equation}
^{\rm I}V^{ij}=Tr((^{\rm I}{\bf P}+^{\rm I}\!{\bf Q}){\bf b}^{ij})
\end{equation}
and
\begin{equation}
^{\rm I}T^{ij}=Tr((^{\rm I}{\bf P}-^{\rm I}\!{\bf Q}){\bf b}^{ij})
\end{equation}
where $(^{\rm I}{\bf P}+^{\rm I}\!{\bf Q})$ is the total charge density matrix and
$(^{\rm I}{\bf P}-^{\rm I}\!{\bf Q})$ is the total spin density matrix for region I.

\section{An example of regional partitioning}
\label{exrp}

To investigate the electronic structure of La$_{2}$CuO$_{4}$, a parent 
compound of high-temperature superconducting materials such as
La$_{2-x}$Sr$_x$CuO$_{4}$, we have performed~\cite{La} extended 
first-principles cluster
calculations. Several clusters containing up to nine copper atoms embedded 
in a background potential were investigated. In Fig.~\ref{homocuo} the 
highest occupied molecular orbital of the cluster
Cu$_9$O$_{42}$/Cu$_{12}$La$_{50}$ is shown. All electron triple-zeta basis 
sets (6-311G Gaussian functions) were used for nine Cu and 42 O atoms resulting
in a total of 663 electrons.

The detailed results of the spin polarized calculations using the local cluster
approximation with generalized gradient corrections (BLYP functional)
will be given in Sec.~\ref{RegionI}. Anticipating these results, here
the contributions to $^{\rm R}G^{ij}_{\alpha}$ and
$^{\rm R}G^{ij}_{\beta}$\ from the three regional partitions R = I, II, and
III, respectively, for
the central copper and the oxygen atom indicated in Fig.~\ref{homocuo}
are collected in Table~\ref{cuoab}.

In contrast to point-ion charge models where only the (small) valence charge 
is considered, the electronic structure is here determined by using
all-electron basis sets (including core electrons) on the atoms in the centre
of the cluster.
The contributions of the nuclear point charges $W$ is, however, 
cancelled to a large extent by the off-site contributions 
$^{\rm III}G_{\alpha ,\beta}$.  Adding and subtracting the contributions 
from the $\alpha$ and $\beta$ spin projections we get the EFG components 
and dipolar hyperfine couplings, respectively, listed in
Table~\ref{cuoabsum}.
It is seen that the combined contributions from region III and W are small. 
For the oxygen, the values from region II give a reduction of the main 
contributions from region I by about 10 \%, for copper by 20 \%. 
For the dipolar hyperfine couplings, the contributions
from regions other than I mostly cancel.

It is remarkable that the calculations on the small clusters 
shown in Fig.~\ref{clusterfig} already give values for the EFG and the dipolar
hyperfine couplings that are close to those obtained from the large 
cluster with nine copper ions. With the cluster CuO$_6$/Cu$_4$La$_{10}$
(Fig.~\ref{clusterfig}(a) where X=Y=La) we obtain at the copper the values
$V_{zz} = 1.396$ and $T_{zz} = -3.526$. With the cluster 
Cu$_2$O$_{11}$/Cu$_6$La$_{16}$ (Fig.~\ref{clusterfig}(b) where X=Y=Z=La)
we get $V_{zz} = 1.167$ and $T_{zz} = -3.467$ and, 
for the O as target ion, $V_{xx} = -0.862$ and $T_{xx} = 0.652$.
This demonstrates that these properties depend on the local
charge and spin distributions and that cluster approaches are 
especially suited for their detailed investigations.


\section{Correlation of first-principles partitions with Sternheimer terms}
\label{fpp}
\subsection{Introductory remarks}

\label{four}
Ideally we would like to have a correspondence between the terms of the
first-principles calculation, Eq.~(\ref{EFGzus}),  
with the semi-empirical Sternheimer equation, Eq.~(\ref{vstern}).  
But the approaches are quite different.  
The first-principles approach  
is a straightforward application of molecular orbital theory  
using the density functional method to obtain the electron density from which  
the expectation value of the appropriate operator (Eq.~(\ref{dijnabla})) is calculated.  
We would like to emphasise that the particular choice of theoretical method
is not crucial. The same argument applies for Hartree-Fock or improved methods
like multi-configuration self-consistent field, M{\o}ller-Plesset and configuration
interaction methods.
The limitations of the approach are largely determined by available computer  
resources which in our case effectively determine the size of the cluster we  
can use.   

In the Sternheimer approach the starting point is to regard the target ion  
in a crystal as isolated and then add terms to compensate for the recognised  
interactions.  Although historically the semi-empirical approach always  
precedes the first-principles methods, sometimes by many years, the value of  
the Sternheimer equation should not be discarded lightly since it appeals to  
the perturbed atomic picture, a model which has been and still is the  
cornerstone of experimental chemistry, rather than the more intractable molecular  
picture.  On the other hand, as in chemistry  where some molecules  
demonstrate high degrees of electron delocalisation, it is necessary to  
abandon the atomic picture in favour of special molecular models.  A good  
example is the separation of aliphatic and aromatic organic chemistries.  
However the subject matter is enormous in both these areas and the  
separation is more than justified.  This is to be contrasted with the  
relatively small number of high-T$_c$ superconducting materials so  
it was natural to pursue the perturbed atom approach. 
It is clear that because of the above differences in approach an immediate  
correspondence between the first principles approach and the semi-empirical  
approach is not to be found.  But since we understand the terms in the  
SE there is a chance of some correspondence. 

\subsection{Identification of terms in the Sternheimer Equation}
\label{identification}

Firstly we look at the $V_{ii}^{lattice}$ term in Eq.~(\ref{vstern}).
This is the contribution to the EFG of the environment of the target ion.  
In Eq.~(\ref{EFGzus}) this is the purely off-site contribution of the
cluster, $^{\rm III}V_{ii}$, and the contribution from the point charges  
surrounding the cluster, $W_{ii}$ (Eq.~(\ref{efgpch})):
\begin{equation}  
V_{ii}^{lattice}=^{\rm III}\!\!V_{ii}+W_{ii} .  
\end{equation} 

$V_{ii}^{local}$ is an atomic (ionic) term which is at the core of the  
semi-empirical perturbation approach.  Cluster calculations are analogous to  
molecular orbital calculations where the properties of atoms largely  
disappear as identifiable entities although some analyses of the  
electron density distribution attempt to allocate charges to particular atoms  
(such as the commonly used Mulliken\cite{mulliken} population analysis).  
However it must  
be stressed that such analyses serve only as a useful guide, and are not  
without controversy.  One study\cite{politzer} concluded that 
the L\"owdin\cite{lowdin}
population analysis was more appropriate than either the 
Mulliken\cite{mulliken} or
a modified Mulliken\cite{politzer} designed for heteronuclear bonds.  

Previous cluster calculations have shown that the degree of overlap and  
delocalisation of the target ion electrons is considerable which 
places in doubt the  
validity of perturbation methods using an isolated ion as the zeroth function. 

$^{\rm I}V_{ii}$ is the closest we can get to $V_{ii}^{local}$ since
$^{\rm I}V_{ii}$ is  
the contribution to $V_{ii}$ from only the basis orbitals centred on the  
target atom.





The semi-empirical approach is well aware of the potential distortion, and  
associated anti-shielding effect on $V_{ii}$, of the target ion by the 
environment of charges.  The $R$ and $\gamma$ terms were introduced to 
accommodate this.  Since the  
first-principles calculation purports to include such distortions  
automatically in $^{\rm I}V_{ii}$ we are inclined
to absorb $-\gamma V_{ii}^{lattice}$ into $V_{ii}^{local}$ replace it in
Eq.~(\ref{vstern}) by~ $^{\rm I}V_{ii}$ to give approximately:

\begin{equation}  
V_{ii}=(1-R) \; ^{\rm I}V_{ii} + ^{\rm III}\!V_{ii} + W_{ii}.
\label{new1}
\end{equation} 

However comparing Eq.~(\ref{new1}) with Eq.~(\ref{EFGzus}), noting that we 
have already identified the first-principles correlation
of $V_{ii}^{lattice}$, we deduce that
\begin{equation}  
-R\; ^{\rm I}V_{ii}= ^{\rm II}\!V_{ii}  
\end{equation}  
enabling us to estimate $R$ from our first-principles regional partitioning
approach.  We obtain (values used in some semi-empirical estimates are in
brackets)
\begin{center}  
$R_{Cu}=0.21~(0.2)~~~{\rm  and}~~~ R_{O_{xx}}=0.084~(0.1)$  
\end{center}  
which seems to justify our approach.  

R was introduced originally to take account of the anti-shielding caused by  
the overlap of charge distributions in the immediate neighbourhood of the  
target atom. 

It should be noted that $R_{O_{yy}}=0.081$ and $R_{O_{zz}}=0.101$ showing 
that $R$
is not necessarily just a simple scalar parameter.  However for the more
symmetrical situated copper ions in the lattice
with an axially symmetric EFG all components $R_{{Cu}_{ii}}$
are identical.

The apparent correspondence of our calculations with the
terms of the Sternheimer equation to obtain $R$ above is unsatisfactory in  
that the quantity $\gamma$ has had to be absorbed into the ``target free ion",
in other words, $^{\rm I}V_{ii}=f(\gamma)$.  Therefore $R$ is a function of $\gamma$,
a result which questions the usefulness of either.  This doubt is reinforced
for $\gamma$ (see Ref.~\cite{abragam}) since the large values necessary 
for the parameter
clearly appear unsuitable to regard the anti-shielding as a perturbation.

However before we disregard $\gamma$ entirely we have calculated it
independently, assuming that the SE equation is correct, using first-principles
calculations on a cluster used to model the effects of doping.
The model and results are described in Appendix~\ref{gamma},
and agree with estimates in the literature that $\gamma$ 
is uncomfortably large.

The evidence here points to the conclusion that the cuprate compounds cannot
be satisfactorily analysed using formulae based on first-oder 
perturbation theory.

\subsection{EFG contributions from region I}
\label{RegionI}
This section discusses the contributions to the EFG from 
the local, on-site term, $^{\rm I}\!M^{ij}_{m,\alpha}(\vec{R}_j)$, first in a 
simplifying approach and then rigorously.   This is important because
V$_{ii}^{local}$ is assumed to be ``exact'' in the Sternheimer equation.

The basis functions at each centre are normally chosen to be radial functions,
multiplied by a spherical harmonic, that is, hydrogen-like. The $s$-functions 
are spherical, the $p$-functions always occur as a group $p_x$, $p_y$ and $p_z$,
the $d$-functions as the group $d_{z^2-r^2/3}$, $d_{zx}$, $d_{yz}$,
$d_{x^2-y^2}$, $d_{xy}$, etc.

A simplifying approach is based on the following argumentation. If all the 
functions in a group are equally occupied the associated electron density 
is spherical and the contribution to the EFG will be zero. So we are only 
interested in those functions which form part of the non-spherical density. So 
an electron configuration in the valence orbital $p^1$ will be analogous to 
the configuration $p^5$ or $p^2$ (if both electrons have the same spin)
that is, equivalent to a single hole in a spherical density. Of course here 
the ion will be in a crystal field and the degeneracies within each group 
may be lifted in which case we can be more definite than saying just $p^1$ 
but, depending on the choice of axes, $p_x^1$.

This simplifies Eq. (\ref{alpha}) to just $^{\rm I}\!M^{ij}_{\eta,\alpha}(\vec{R}_J)$
where $\eta$ is the orbital which introduces the asymmetry into the spherical
distribution about centre $J$.  It is possible that the asymmetry is caused by the $\beta$-
spin orbital so the term we should use $^{\rm I}\!M^{ij}_{\eta,\beta}(\vec{R}_J)$ as would
be case in $p^5$ for example, a $\beta$ hole in the spherical distribution.

Even $^{\rm I}\!M^{ij}_{\eta,\alpha}(\vec{R}_J)$ is simplified because in this discussion only one
orbital on centre $J$ is involved which we denote by $g$
\begin{equation}
^{\rm I}G^{ij}_\alpha=\,^{\rm I}\!M^{ij}_{\eta,\alpha}(\vec{R}_J)=
c_{\eta,\alpha}^{J,g}c_{\eta,\alpha}^{J,g}D^{ij}_{J,g,J,g}(J).\label{oneorbital}
\end{equation}

This is therefore the required result subject to all these approximations introduced above
\begin{equation}
V_{ij}(\vec{R}_J)=\,^{\rm I}G^{ij}_\alpha\propto n_g<r^{-3}>_g\label{final}
\end{equation}
where $n_g$ is the occupancy of orbital $g$.  In particular
\begin{equation}
V_{xx}(O)=\frac{2}{5}[2 \times n (2p_x) - n (2 p_y) -n(2 p_z)] < r^{-3}>_{2p}
\label{vxxo}
\end{equation}
for the oxygen and
\begin{eqnarray}
V_{zz}(Cu) & = &\frac{2}{7}[2 \times n (3d_{x^2-y^2}) + 2 \times n (3d_{xy}) 
\nonumber \\
& - & n (3d_{zx}) - n (3d_{yz}) - 2 \times n (3d_{3 z^2 - r^2})] 
< r^{-3}>_{3d} \label{vzzcu}
\end{eqnarray}
for the copper, and similarly for the other components of $V_{ii}$.

The right hand side of these equations (\ref{vxxo},\ref{vzzcu}) is just the 
form of the equation often used to estimate the occupancies, $n$, of orbitals.  
$<r^{-3}>_g$ is taken as the value for the atom or ion from calculation.  
This may be a poor estimate since $<r^{-3}>$ is expected to be affected 
significantly by the crystal field which will tend to lift degeneracies 
and concentrate the asymmetries along particular directions.

In the rigorous approach, the on-site matrix elements 
$^{\rm I}\!M^{ij}_{m,\alpha}$ are determined in the following way. The 
expectation values $< r^{-3}>_{m,\alpha}$ are performed analytically. The
coefficients $c_{\gamma,\alpha}^{J,g}$ are given by the $\gamma,\alpha$
eigenvectors of the selfconsistent field equations. It must be emphasised 
that in the basis set 6-311G there are three radial functions for each 
of the five $3d$ orbitals of Cu and three for each of the $2p$ orbitals.
We remark that the relation
\begin{equation}
\frac{4}{5} \times c_{m,\alpha}^{J,m}c_{m,\alpha}^{J,m} \times 
<r^{-3}>_{m,\alpha} = \, ^{\rm I}\!M^{xx}_{m,\alpha}(\vec{R}_J) 
\label{rm3mxx}
\end{equation}
which is valid for an orbital $m$ showing $2p_x$-symmetry implies the relation
\begin{equation}
\frac{4}{5} \times \,^{\rm I}\!N_{\alpha}(2p_x) \times 
\,^{\rm I}\!\!<r^{-3}>_{\alpha} = \,^{\rm I}G^{xx}_{\alpha}
\label{npalpha}
\end{equation}
after performing the average over all the orbitals having the same symmetry. 
The resulting values for the oxygen in the above mentioned Cu$_9$O$_{42}$ 
cluster are given in Tables~\ref{pIa} and \ref{nI}.

We first note that the expectation values $<r^{-3}>$ differ for the three 
p-orbitals and the spin projections by several percents in contrast to the 
assumptions in the simplified approach. It is evident that $<r^{-3}>_x$
differs from the other components since the bonding is along the x-direction,
but also $<r^{-3}>_y$ and  $<r^{-3}>_z$ differ. This is due to a nonsymmetric 
distribution of the electron densities as it is plotted in Fig.~\ref{denasyo}
for the $2p$ electrons in oxygen.
Furthermore the values $<r^{-3}> \sim 4$ are about 10 \% larger than those
obtained from calculations on isolated atoms or ions.
In Table~\ref{nI} we have also given the Mulliken partial charges
(see Eq.~(\ref{mulli})) which show that the two 2$p_{\pi}$ are effectively
fully occupied. It is only the 2$p_{\sigma}$ AO which is involved in the 
bonding and the convey of spin density from the copper to the ligand.

The results of the analogous analysis of the contributions to the EFG and
hyperfine dipole tensor for the Cu target ion are collected in 
Tables~\ref{cupIa} and \ref{cunI}. The values called ``remainder(s,d)''
come from terms in the evaluation of matrix elements~(\ref{onecentre}) where 
one basis function is s-like and the other d-like.
The contributions from the Cu $p$-type orbitals to the EFG are substantial.
This is due to the large  $< r^{-3} >$ values. The occupancies of these 
orbitals, however, are close to one. The same applies to the three d
orbitals with $t_{2g}$ symmetry as is seen from the Mulliken partial 
charges in Table~\ref{cunI}.
As expected, the distinguished AO is the $3d_{x^2-y^2}$ accompanied with
some polarization in the $3d_{3z^2-r^2}$. 

Note, again, that $<r^{-3}>_{3d}\; \approx 8$, in contrast to the 
value $\approx 6$ used in approximate procedures.

In a similar way the contributions $^{\rm II}G^{ij}_{\alpha}$ and
$^{\rm II}G^{ij}_{\beta}$ from region II can be analyzed. The results
are given in Appendix~\ref{RII}.

\subsection{Electron-hole symmetry}
\label{ehs}
It should be emphasized that the occupations $^{\rm I}\!N_{\alpha}$ are
determined by the expansion coefficients of the occupied MO into the 
individual AOs. To connect the results of the {\it ab initio} calculations
with EFG analyses using the hole picture we can identify the unoccupied 
MOs which lie lowest in energy as contributions from ``holes". In 
particular, in our example of a Cu$_9$O$_{42}$ cluster, there are nine such
unoccupied MOs above the highest occupied molecular orbital (HOMO) 
which all show predominantly $3d_{x^2-y^2}$
and $2p_{\sigma}$ character on the copper and oxygen rows, respectively.
If we assume that they were occupied we would get contributions to the
EFG and hyperfine tensor which we define by $\overline{V}_{ij}$ and 
$\overline{T}_{ij}$. These are collected in Tables~\ref{efghypop} (for
oxygen) and \ref{efghypcu} (for copper) together with the values $V_{ij}$
and $T_{ij}$ as calculated from all occupied MOs (see Table~\ref{cuoab}).

For an isolated ion, one has the relations
\begin{equation}
V_{ij} + \overline{V}_{ij} = 0
\label{vvover}
\end{equation}
and
\begin{equation}
T_{ij} + \overline{T}_{ij} = 0.
\label{ttover}
\end{equation}
For ions in the cluster, Eq.~(\ref{vvover}) is not necessarily correct
since the environment is generally non-spherical as is shown in
Sec.~\ref{DMA}.
Tables~\ref{efghypop} and \ref{efghypcu}
show that relation (\ref{ttover}) approximately holds but that (\ref{vvover})
is not fulfilled.
In this respect we remark that a calculation on the cluster (CuO$_6$)$^{-11}$ 
(see Fig.~\ref{clusterfig}) where the Cu is nominally in a $d^{10}$ state, 
yields $V_{zz} = -2.412$.

\subsection{Density matrix argument}
\label{DMA}

Since, contrary to the usual assumption, 
Eq.~(\ref{vvover}) is  not correct under all conditions
we will present the detailed theoretical background in this Section.
We will not initially refer to the electron spin projection and so the terminology 
will be identical to that used in Sec.~\ref{dmf} save for the $\alpha$ subscript.

So from Sec.~\ref{dmf} Eq.~(\ref{dm1})
\begin{equation}
\Phi= {\bf c}^\dagger{\bf B}
\end{equation}
where ${\bf c}$ is the matrix of MO coefficients collected in columns.
The overlap matrix is defined as (using $\cal{O}$=1 in Eq.~(\ref{dm2}))
\begin{equation}
{\bf S}={\bf b}^1 = <{\bf B}|{\bf B}^\dagger >.
\end{equation}
However it is more convenient for us to use an orthogonal, but entirely equivalent,
basis set of orbitals.  We will label these by the column matrix ${\bf B}^\prime$
such that
\begin{equation}
{\bf B}^\prime= {\bf S}^{-\frac{1}{2}}{\bf B}
\end{equation}
and
\begin{equation}
{\bf I}_{n_c}=<{\bf B}^\prime|{\bf B}^{\prime \dagger} >
\end{equation}
where ${\bf I}_{n_c}$ is the unit matrix of dimension $n_c$.  
The MO's are now written as
\begin{equation}
\Phi= {\bf d}{\bf B}^\prime
\end{equation}
where ${\bf d}={\bf c}{\bf S}^{\frac{1}{2}}$, the MO coefficients in the orthogonal basis.
We now define spin MO's ${\bf \Phi}_\alpha$,
to hold electrons with spin projection $+\frac{1}{2}$, and
${\bf \Phi}_\beta$,
to hold electrons with spin projection $-\frac{1}{2}$.
Since each spin orbital can only hold one electron we need two
density matrices, {\bf P} and {\bf Q}, to describe the $\alpha$-spin and $\beta$-spin
densities respectively.
\begin{equation}
{\bf P}={\bf d}_\alpha{\bf I}_\alpha{\bf d}_\alpha^\dagger~~~~~~~~~~~~~{\bf Q}={\bf d}_\beta{\bf I}_\beta{\bf d}_\beta^\dagger
\end{equation}
where ${\bf I}_\alpha$ is a diagonal $m\times m$ matrix with 1's for each
occupied $\alpha$-spin and zeroes otherwise.  ${\bf I}_\beta$ is similar.
The charge density matrix, necessary to calculate the EFGs, is given by
${\bf P}+{\bf Q}$ and the spin density matrix, necessary to calculate the
hyperfine tensor, is given by ${\bf P}-{\bf Q}$.  In keeping with previous practice
we can evaluate the ``hole" density.  The ``hole" density is simply the total
empty Hilbert space ${\bf I}_m$ minus the {\bf P} or {\bf Q}.  The charge 
``hole" density is
\begin{equation}
{\bf I}_m - {\bf P} + {\bf I}_m - {\bf Q} =2{\bf I}_m -({\bf P} + {\bf Q})
\end{equation}
and the spin ``hole" density is
\begin{equation}
{\bf I}_m - {\bf P} - ({\bf I}_m - {\bf Q}) = -({\bf P} - {\bf Q}).
\end{equation}
Since the former includes the diagonal matrix 2${\bf I}_m$ this can make a
contribution to the ``hole" EFG calculation.  If the Hilbert space is not spherical,
or at least does not possess cubic symmetry, the contribution will be non-zero.
So the relation 
given in equation~\ref{vvover} is not strictly valid in ions
where the degeneracy of the $d$-type orbitals is lifted.  The degeneracy
is only lifted if the symmetry is less than cubic.  In all cases of practical interest
\begin{equation}
V_{ij} + \overline{V}_{ij} \neq  0 .
\end{equation}
On the other hand the spin ``hole" density is simply the negative of the
spin density matrix leading to a verification of Eq.~(\ref{ttover})
\begin{equation}
T_{ij} + \overline{T}_{ij} = 0.
\end{equation}

We can use this argument to explain why, in Tables \ref{efghypop} and \ref{efghypcu},
the difference between the EFG's calculated from the occupied orbitals
and the EFG's approximately calculated from selected unoccupied orbitals
differ more markedly than the difference between equivalent calculations
for the hyperfine tensors.  If all the unoccupied orbitals are taken there is
no difference between the occupied calculation and unoccupied calculation
for the hyperfine tensor.


\section{Discussion and Conclusions}
\label{summary}

The number of problems which can be solved exactly by wave mechanics
is very small and the perturbation method, originally devised for classical
systems, was developed.  Essentially, used in its less rigorous form, the
problem is reduced to identifying that part which is well understood and
treating the rest as a perturbation.  In ionic crystals where
the properties, for example the EFG, of one of the ions (the target ion)
is of interest, the purpose is to try and predict how changes to the
properties of the isolated target ion can be accounted for by
perturbation from its environment.  In the mathematics of
perturbation theory the changes to the target ion
wave function can be achieved by mixing in the excited states of the
target ion.  Since the excited states form a complete set of functions this is
always true although probably a very inefficient process.

An environment of ions (point charges) contributes to the EFG at the target
ion but will also interact with the electrons of the target ion to cause
a distortion which in turn changes the electron contribution to the EFG.
Such distortions should be easily simulated by judiciously mixing
in the excited states of the target ion with its ground state.  However
if the possibility of covalent bonding occurs two problems seem to arise.
Firstly the overlap of orbitals with nearest neighbour ions to the target ion
and secondly the possibility, in a Mulliken population sense, of a transfer
of electronic charge.  Although in principle this can be accommodated by
including the excited states of the target ion it is hardly a small
perturbation questioning the applicability of the method to crystals
where the possibility that covalent bonding occurs.

The Sternheimer equation (Sec.~\ref{st}) uses a first-order perturbation 
theory type argument to obtain the individual terms which we attempt 
to correlate with different regions of the crystals (see Secs.~\ref{rp} 
and~\ref{dmf}). Starting from a lattice of ionic charges the contribution 
to the EFG can be easily calculated at the target ion (Sec.~\ref{hfc}).
Of course the target ion, assumed to be a point charge makes no contribution.
The electronic structure about the target ion is very important (Secs.~\ref{ehs}
and ~\ref{DMA}) as long as it is not spherically symmetric. Since the crystal 
lattice interacts with the target ion (a crystal field) any asymmetry in the 
electron distribution (for example unfilled shells) will be significant since 
the crystal field will lift some degeneracies. The $V_{ii}^{local}$ term is 
therefore crucial and fortunately is easily amenable to accurate calculation 
and transferable for the same ion to other crystals with a different chemical 
constitution. Although this crystal field could be guessed as being a small 
perturbation this is clearly not supported by the large value of $\gamma$ 
calculated here which are the same order of magnitude as those obtained 
``experimentally'' (see also Appendix~\ref{RII}).

However the shielding parameter R (Sec.~\ref{identification}), intended to 
take account of ``overlap'' with nearest neighbour ions is rather more 
difficult to justify as a first order perturbation parameter. The ``overlap'' 
with the nearest neighbour orbitals could potentially lead to large electron
density distortions, particularly of the outer shells, and also significant
charge transfer. A better representation of this intuitive picture is the
multi-center model which is at the core of molecular orbital theory.
Unfortunately this complicates the interpretation of R and reduces its 
usefulness as does the conclusion that R and $\gamma$ are not independent.

The use of experimental or theoretical data from isolated ions has long
been a method of extracting information from a crystal system.  Even without
the complications of crystal fields or ``overlap'' the very existence of the
lattice surrounding the target ion produces an unyielding restrictive cage
from Pauli's exclusion principle.  Ions can be attributed an ionic radius
which apparently determines the structures of many ionic crystals.  Any
transfer of electronic charge onto the ion will hardly be able to use
valence orbitals of the expected ``free ion'' size.  This will no doubt
contract the inner shell orbitals to compensate changing the experimental
``free ion'' EFGs.

We conclude that the perturbed ion approach which results in the
Sternheimer equation is inappropriate for cuprate crystals which are
common in high-temperature superconducting materials mainly due to the
significant covalent bonding in the CuO$_2$ planes.  However this has
other consequences since the perturbation model also suggests an easy
method to estimate the ``holes'' in the electron structure from the EFGs,
whose distribution in turn is essential for models of 
superconductivity itself. We have shown that these estimates are 
probably wrong and at the very least
their values should be reassessed.
Therefore precise information on the charge and spin density distributions in
copper oxides is necessary and EFGs, determined by nuclear
quadrupole resonance spectroscopy, can help to provide this information.
It is necessary, however, that they are analysed in a more sophisticated
manner than with point-charge models.

\acknowledgments
We acknowledge the help of P. H\"usser, M. Mali, S. Pliber\v{s}ek, S. Renold, 
J. Roos and H. U. Suter. 
In particular, we would like to express our gratitude to R. E. Walstedt for 
enlightening discussions. 
This work has been supported by the Swiss National Science Foundation.
A major part of the computation was carried out at the national supercomputing 
center CSCS.
\appendix
\section{Contact densities}\label{contactdensities}
The regional partitioning for the evaluation of the EFG tensors and 
the dipolar hyperfine tensors applies also for the contact interaction.
The corresponding results are given in this appendix.

We denote the contact density for the target nucleus $J$ as
\begin{equation}
D(\vec{R}_J) = \frac{8 \pi}{3} \left( \sum_m \mid \psi_m^{\uparrow}(\vec{R}_J) \mid^2
                 - \sum_{m'} \mid \psi_{m'}^{\downarrow}(\vec{R}_J) \mid^2 \right) 
\end{equation}
where the sum extends over the occupied MOs and perform the same regional partitioning as 
in Sec.~\ref{exrp}. The total contributions to $D_{ns}({\rm Cu})$ and $D_{ns}({\rm O})$
for the different s-like AOs are listed in Table~\ref{FC} with the small 
contributions from regions II and III given in parentheses. Since the expectation values 
$\mid\psi_{ns}(\vec{R}_J)\mid^2$ have nearly the same values for spin up and 
down projections we can describe the results also in terms of partial 
polarizations $f_{ns}$ according to
\begin{equation}
D_{ns}(\vec{R}_J) = \frac{8 \pi}{3} \mid \psi_{ns}(\vec{R}_J) \mid^2
                      f_{ns} .
\end{equation}

Note that these results refer to maximal spin-multiplicity. Thus,
the values for $D({\rm Cu})$ and  $D({\rm O})$ include the transferred hyperfine fields from the
four and two nearest neighbour copper ions, respectively. These transferred
hyperfine fields have been discussed extensively in Refs.~\cite{Y,Leipzig}.

\section{Estimate of $\gamma$ parameter}\label{gamma}

We have performed several cluster calculations where
point charges $q$ have been added to the La$^{3+}$ pseudopotentials at 
positions
X, Y, and Z for the clusters CuO$_{6}$/Cu$_4$La$_{10}$ and
Cu$_2$O$_{11}$/Cu$_6$La$_{16}$ (see Fig.~\ref{clusterfig}).
For the cluster CuO$_{6}$/Cu$_4$La$_{10}$ the target ions are the central Cu
and the planar O on the x-axis whereas for Cu$_2$O$_{11}$/Cu$_6$La$_{16}$ 
the target ions are the Cu to the right and the central O.

Since these additional charges are in region III, 
the differences in the calculated EFG tensors are then 
identified with the term $\Delta V_{ii}(q) = (1 - \gamma ) \Delta W(q)$.

The results are collected in Tables~\ref{gamma1} and \ref{gamma2}. 
It is seen that these $\gamma$ values are unreasonably large and that these
``lattice'' contributions in the SE cannot be used at all. What really
happens is that the additional charges distort and polarize the nearby ions
(oxygens in the present cases) which in turn then influence the target ion.

\section{Contributions from region II}\label{RII}

For completeness we collect here the results of the analysis of the 
contributions from region II (see Sec.~\ref{RegionI}). With the oxygen as target 
nucleus, the values of $^{{\rm II}}G_{\alpha}^{ii}$ and
$^{{\rm II}}G_{\beta}^{ii}$ are given in Table~\ref{pIIa}.
Note that the contributions assigned to s-character are due to matrix 
elements of the operator $\cal{D}$ between s-type functions centred at the oxygen and
d-type functions centred at the neighboring copper nuclei.

In Table~\ref{cupIIa} the contributions from 
region II for the copper target nucleus are given.

  

\newpage
\begin{table}
\centering
\caption{Diagonal elements of the tensors $G$ from the contributions of on-site
(I), on-site/off-site (II) and off-site AOs (III) for spin 
projections $\alpha$ and $\beta$ for Cu and O. Contributions from nuclear 
charges $W$, EFG tensor $V$ and hyperfine tensor $T$ .
}
\baselineskip=24pt
\vspace{5mm}
\begin{tabular}{|llrrrlrrr|}
\hline
  &  &  &  Cu  &  &  &  &  O  &  \\
 $G$ &  & ${xx}$ & ${yy}$ & ${zz}$ &    & ${xx}$ & ${yy}$    &   ${zz}$  \\
\hline
$^{\rm I}G_{\alpha}$ & & $ 0.432$ & 0.432  & $-0.864$& &$-0.141$ & 0.130  & 0.011 \\
$^{\rm I}G_{\beta}$& & $-1.266$ &$-1.266$& 2.532& &$-0.787$ & 0.436  & 0.351  \\
\hline
$^{\rm II}G_{\alpha}$& &0.073 & $ 0.073$ & 
$-0.146$& &0.018 & $-0.013$ & $-0.005$ \\
$^{\rm II}G_{\beta}$ & & 0.102 & $ 0.102$ & 
$-0.204$& &0.060& $-0.033$ & $-0.027$ \\
\hline
$^{\rm III}G_{\alpha}$& &0.278 & $ 0.278$ & $-0.556$& &1.367 & $-0.418$ & $-0.949$\\
$^{\rm III}G_{\beta}$ & & 0.262 & $ 0.262$ & $-0.524$& &1.286& $-0.390$ &$-0.896$\\
\hline
$W$ & &$ -0.522$& $-0.522 $& $ 1.044$ & & $-2.693$& 0.832 &$ 1.861 $ \\
\hline
$V$ & & $-0.642$ &$-0.642$ & 1.283 & & $-0.890$ & 0.545 & 0.345\\
$T$ & & 1.685 & $ 1.685$ & $-3.370$ & & 0.685 & $-0.314$ & $-0.371$\\
\hline
\end{tabular}
\label{cuoab}
\end{table}

\begin{table}
\centering
\caption{Contributions to the EFG and the hyperfine coupling tensor 
from the different regional partitions.
}
\baselineskip=24pt
\vspace{5mm}
\begin{tabular}{|rrrrrrrrr|}
\hline
Region &$V_{zz}$(Cu) &$T_{zz}$(Cu)&$V_{xx}$(O)&$V_{yy}$(O)&$V_{zz}$(O)
       &$T_{xx}$(O)&$T_{yy}$(O)&$T_{zz}$(O) \\
\hline
 I       & 1.668       & $-3.396$   & $-0.928$ & 0.566 & 0.362 & 0.646 & $-0.306$ & $-0.340$\\
 II      &$-0.350$     & 0.058      & 0.078    &$-0.046$&$-0.032$& $-0.042$& 0.020 & 0.022 \\
 III + W  &$-0.036$     & $-0.032$  & $-0.040$ &0.024&0.016 & 0.081&$-0.028$&$-0.053$  \\\hline
Total    & 1.282       & $-3.370$   & $-0.890$ &0.544 & 0.346 & 0.685 &$-0.314$&$-0.371$ \\
\hline
\end{tabular}
\label{cuoabsum}
\end{table}

\begin{table}[ht,b]
\centering
\caption{Contributions of on-site AOs (region I) for spin projection $\alpha$
and $\beta$ for the planar oxygen.
}
\baselineskip=24pt
\vspace{5mm}
\begin{tabular}{|lcrrrcrrr|}
\hline
 & & & $\alpha$ & & & & $\beta$ & \\
          &   & ${xx}$    &  ${yy}$    &   ${zz}$&      &
${xx}$    &  ${yy}$    &   ${zz}$   \\
 \hline
 $p_x$        & & 3.046    & $-1.523$  &  $-1.523$ &&
 2.370    & $-1.185$  & $-1.185$  \\
 $p_y$        & & $-1.613$ & 3.226     &  $-1.613$ &&
 $-1.592$ & 3.184    & $-1.592$ \\
 $p_z$        & & $-1.574$ & $-1.573$  & 3.147 &&
$-1.564$ & $-1.564$  & 3.128    \\\hline
 $^{\rm I}G$&& $-0.141$ & 0.130 & 0.011 &&
$-0.787$ & 0.436 & 0.351    \\
\hline
\end{tabular}
\label{pIa}
\end{table}

\begin{table}[ht,b] \centering
\caption[dummy]{Occupations $^{\rm I}\!N(2p_j)$ and Mulliken partial
charges $\rho$ of the $2p$ orbitals, and averaged values of $<~r^{-3}~>$.
}
\baselineskip=24pt
\vspace{5mm}
 \begin{tabular}{|lrrr|}
\hline
        & $2p_{x}$    &  $ 2p_{y}$    &   $2p_{z}$   \\
 \hline
 $^{\rm I}\!N_{\alpha}$  & 0.921  & 1.018  & 1.006 \\
 $^{\rm I}\!N_{\beta}$   & 0.737  & 1.013  & 1.006  \\\hline
 $\rho_{\alpha}$         & 0.914  & 0.999  & 1.001  \\
 $\rho_{\beta}$          & 0.783  & 0.994  & 1.001  \\\hline
 $^{\rm I}\!< r^{-3} >_{\alpha}$& 4.135  & 3.961 & 3.911 \\
$^{\rm I}\!< r^{-3} >_{\beta}$  & 4.044 & 3.933 & 3.885 \\
\hline
 \end{tabular}
\label{nI}
\end{table}

 \begin{table}[ht,b]
\centering
\caption{Contributions of on-site AOs (region I) for spin projections $\alpha$
and $\beta$ for the central copper.
}
\baselineskip=24pt
\vspace{5mm}
\begin{tabular}{|lcrcr|}
\hline
$^{\rm I}G_{zz}$ &&$\alpha$ && $\beta$ \\
 \hline
Remainder(d,s)       && $0.214 $ &&  0.232  \\
$p$      & & $-0.651 $ && $-0.599 $ \\
 $d_{x^2-y^2}$  &&  $-4.839 $ && $-1.325 $ \\
 $d_{z^2-r^2/3}$&&  $ 4.453 $ && $ 4.277 $ \\
 $d_{xy}$       && $-4.563 $ && $-4.538 $ \\
 $d_{zx}$, $d_{yz}$  && $4.522 $ && $4.485 $ \\\hline
 $^{\rm I}G_{zz}$ && $-0.864 $ && 2.532  \\
\hline
\end{tabular}
\label{cupIa}
\end{table}
\begin{table}[ht,b] \centering
\caption[dummy]{Occupations $^{\rm I}\!N$ and Mulliken partial charges $\rho$
of the $3d$ orbitals, and averaged values of $< r^{-3} >$.
}
\baselineskip=24pt
\vspace{5mm}
 \begin{tabular}{|lrrrrr|}
\hline
  & $3d_{x^2 -y^2}$ & $3d_{z^2-r^2/3}$ & $3d_{xy}$ & $ 3d_{zx}$ & $3d_{yz}$ \\
 \hline
 $^{\rm I}\!N_{\alpha}$ & 1.030  & 0.966  & 1.000  & 0.998  & 0.998   \\
 $^{\rm I}\!N_{\beta}$  & 0.297  & 0.939  & 1.000  & 0.998  & 0.998   \\\hline
 $\rho_{\alpha}$        & 0.999  & 0.971  & 0.997  & 0.997  & 0.997   \\
 $\rho_{\beta}$         & 0.358  & 0.946  & 0.997  & 0.997  & 0.997   \\\hline
 $^{\rm I}\!< r^{-3} >_\alpha$& 8.224 & 8.066 & 7.982 & 7.926 & 7.926 \\
 $^{\rm I}\!< r^{-3} >_\beta$ & 7.821 & 7.976 & 7.939 & 7.864 & 7.864 \\
\hline
 \end{tabular}
\label{cunI}
\end{table}

\begin{table}[ht,b]
\centering
\caption{Electric field gradients $V_{ii}$ and the hyperfine tensors
$T_{ii}$ for the occupied states and $\overline{V}_{ii}$ and 
$\overline{T}_{ii}$ from the unoccupied states close
but above the Fermi energy for the planar oxygen atom.
}
\baselineskip=24pt
\vspace{5mm}
\begin{tabular}{|lrrrr|}
\hline
 $i$ &  $V_{ii}$ & $\overline{V}_{ii}$ &  $T_{ii}$ & $\overline{T}_{ii}$ \\
\hline
 $x$ &$-0.873$  & $ 0.624$   & 0.685 & $-0.624$  \\
 $y$ &$ 0.563$  & $-0.284$   & $-0.314$ & $0.284$  \\
 $z$ &$ 0.310$  & $-0.340$   & $-0.371$ & $0.340$ \\
\hline
\end{tabular}
\label{efghypop}
\end{table}
\begin{table}[ht,b]
\centering
\caption{Electric field gradients $V_{ii}$ and the hyperfine tensors
$T_{ii}$ for the occupied states and $\overline{V}_{ii}$ and 
$\overline{T}_{ii}$ from the unoccupied states close
but above the Fermi energy for the central copper nucleus.
}
\baselineskip=24pt
\vspace{5mm}
\begin{tabular}{|lrrrr|}
\hline
 $i$ &  $V_{ii}$ & $\overline{V}_{ii}$ &  $T_{ii}$ & $\overline{T}_{ii}$ \\
\hline
 $x,y$ &$-0.623$  & $ 1.670$   & 1.685 & $-1.670$  \\
 $z$ &$ 1.246$  & $-3.340$   & $-3.370$ & $3.340$ \\
\hline
\end{tabular}
\label{efghypcu}
\end{table}

\begin{table}[ht,b]
\centering
\caption{Expectation values of the s-like AOs, $|\psi_{ns}|^2$, 
contact densities $D_{ns}$, and polarizations $f_{ns}$ at Cu 
and O, respectively. For $D_{ns}$ the total values are 
given with the contributions from regions II and III in parentheses.}
\begin{center}
\begin{tabular}{|ccccccc|}
\hline
$n$ & $|\psi_{ns}(\textrm{Cu})|^2$ & $D_{ns}(\textrm{Cu})$ & $f_{ns}[\%]$
& $|\psi_{ns}(\textrm{O})|^2$ & $D_{ns}(\textrm{O})$ & $f_{ns}[\%]$ \\
1 & 7300 & $-0.053\;(0.000)$ & $-8.66 \times 10^{-5}$ &  141 & $-0.262\;(0.001) $ & $-0.0222$ \\
2 & 725  & $-3.637\;(0.002)$  & $-5.99 \times 10^{-2}$ &  6.77& $ 1.516\;(0.055)$  & $ 2.673$ \\
3 & 107  & $ 2.575\;(0.013)$   &   0.288                &  & &     \\
4 & 2.35 & $ 2.037\;(-0.017)$  & $10.35$                &  & &     \\ 
\hline
\end{tabular}
\end{center}
\label{FC}
\end{table}
\begin{table}[ht,b]
\centering
\caption{Contributions to $^{{\rm II}}G$ from region II for spin
projection $\alpha $ and $\beta$
for the planar oxygen.
}
\baselineskip=24pt
\vspace{5mm}
\begin{tabular}{|lcrrrcrrr|}
\hline
      &   &    &   $\alpha$    &   &  &   & $\beta$ &  \\
      &  & ${xx}$    &  ${yy}$    &   ${zz}$
      &  & ${xx}$    &  ${yy}$    &   ${zz}$ \\
 \hline
 $s$         &  &$-0.004 $  & 0.001   & 0.003 
             &  &$0.002 $  & $-0.002 $   & $-0.001 $ \\
 $p_x$       &  & 0.021     & $-0.011 $  &  $-0.010 $
             &  &$ 0.057 $ & $-0.028 $   & $-0.028 $ \\
 $p_y$       &  & $ 0.001 $ & $-0.004 $  &  0.003 
             &  &$ 0.001 $ & $-0.004 $   &  0.003   \\
 $p_z$       &  & $ 0.000 $ & $ 0.001 $  & $-0.001 $
             &  &$ 0.000 $ & $ 0.001 $  & $-0.001 $ \\\hline
 $^{{\rm II}}G$ & & $0.018 $ &$-0.013 $ & $-0.005 $
             &  & $0.060 $ &$-0.033 $ & $-0.027 $  \\
\hline
\end{tabular}
\label{pIIa}
\end{table}
\begin{table}[ht,b]
\centering
\caption{Contributions to $^{{\rm II}}G$ from region II for spin 
projections $\alpha$ and $\beta$ for the central copper.
}
\baselineskip=24pt
\vspace{5mm}
\begin{tabular}{|lcrcr|}
\hline
 $^{\rm II}G_{zz}$ &&$\alpha$ 
&&$\beta$ \\
 \hline
Remainder       &&  $ 0.000 $ &&  0.000  \\
 $s  $          &&  $ 0.001 $ && $ 0.003 $ \\
 $p$          && $ 0.011 $ &&  $ 0.011 $ \\
 $d_{x^2-y^2}$  && $ 0.013 $ &&  $-0.025 $ \\
 $d_{z^2-r^2/3}$&& $-0.172 $ &&  $-0.194 $ \\
 $d_{xy}$       && $ 0.001 $ &&  $ 0.001 $ \\
 $d_{zx}$,  $d_{yz}$   &&$0.000 $ && $0.000 $ \\\hline
 $^{\rm II}G_{zz}$ && $-0.146 $ && $-0.204 $ \\
\hline
\end{tabular}
\label{cupIIa}
\end{table}
\begin{table}[ht] 
\centering
\caption{Calculated values of $\gamma$ for the cluster in 
Fig.~\ref{clusterfig}(a) with additional point charges at positions X and Y.}
\baselineskip=24pt
\vspace{5mm}
\begin{tabular}{|rrrr|}
\hline
  X & Y & $\gamma(Cu)$  &  $\gamma(O)$  \\
 \hline
 $-0.1$         & $-0.1$  & $-51.1$   & $+19$  \\
 $-0.1$         & $0$     & $-51.3$   & $+20$ \\
 $-1.0$         & $0$     & $-50.9$   & +19  \\
 $-1.0$         & $-1.0$  & $-52.1$   & $+23$  \\
\hline
\end{tabular}
\label{gamma1}
\end{table}
\begin{table}[ht] 
\centering
\caption{Calculated values of $\gamma$ for the cluster in 
Fig.~\ref{clusterfig}(b) with additional point charges at positions X, Y, and 
Z.
}
\baselineskip=24pt
\vspace{5mm}
\begin{tabular}{|rrrrr|}
\hline
 X & Y & Z & $\gamma(Cu)$  &  $\gamma(O)$  \\
 \hline
 $-0.1$         & $-0.1$  & $-0.1$  & $-37$   & $-33$  \\
 $0$            & $-0.1$  & $-0.1$  & $-40$   & $-32$  \\
 $0$            & $0$     & $-1.0$  & $-26$   & $-33$  \\
\hline
\end{tabular}
\label{gamma2}
\end{table}
\newpage
\begin{figure}
{\centerline{\includegraphics[height=6cm]{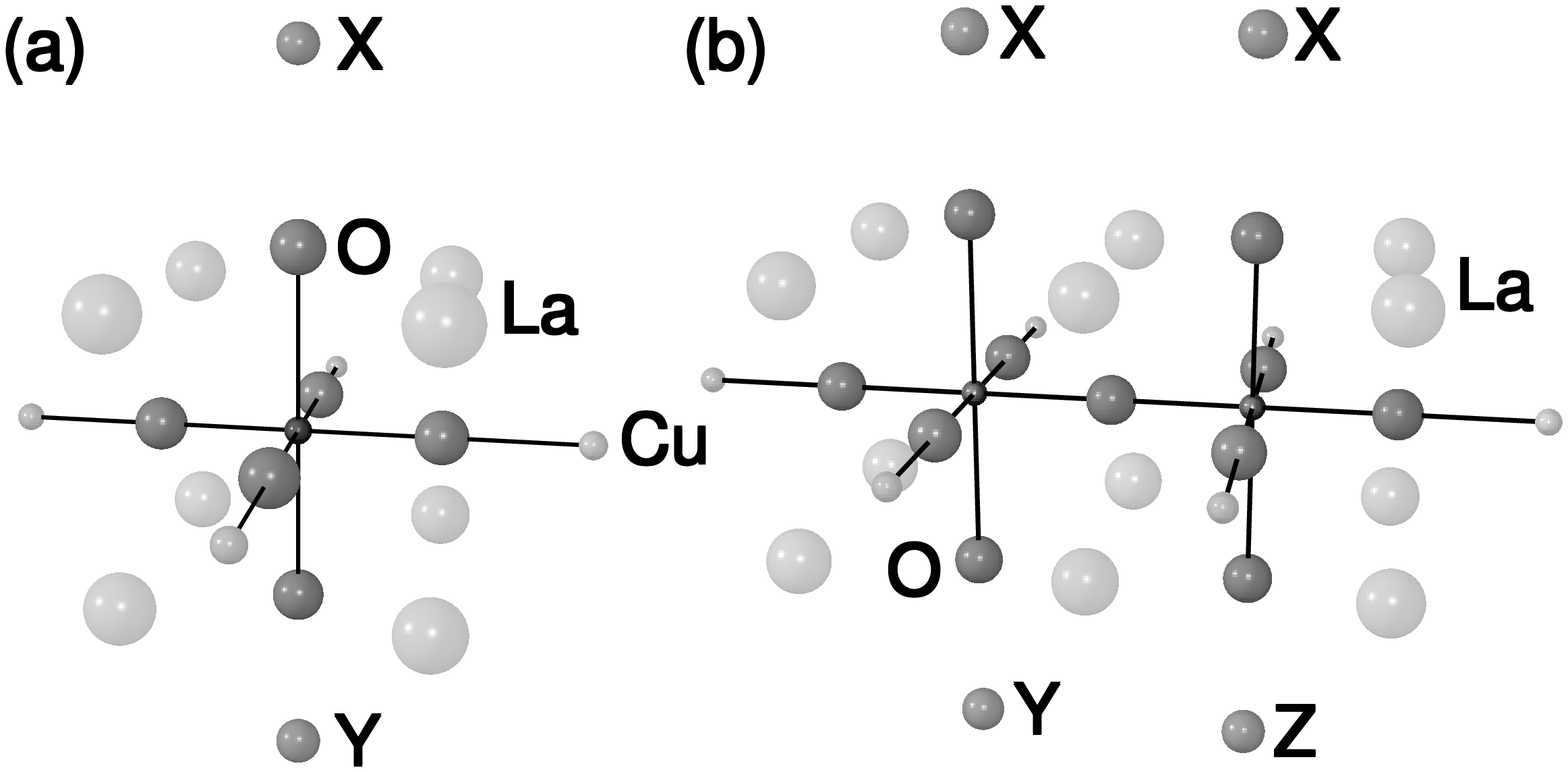}}
\caption{\label{clusterfig} The CuO$_{6}$/Cu$_4$La$_{10}$ and
Cu$_2$O$_{11}$/Cu$_6$La$_{16}$ clusters. The notations X, Y, and Z are for
later reference (see Sec. Appendix \protect{\ref{gamma}}).}}
\end{figure}

\begin{figure}
{\centerline{\includegraphics[height=6cm]{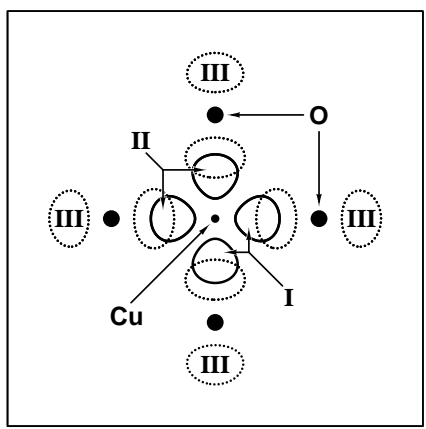}}
\caption {
Illustration of the contributions I to III to expectation values in a
CuO$_4$ cluster. The $J$th atom is the central Cu, whereas the neighboring
O-atoms denote the atoms $K$ and $L$. The full curve limits the $d$-electrons
of the central Cu and the dotted curves enclose the oxygen $p$-electrons.
\label{CudOp}}}
\end{figure}

\begin{figure}
{\centerline{\includegraphics[height=6.2cm]{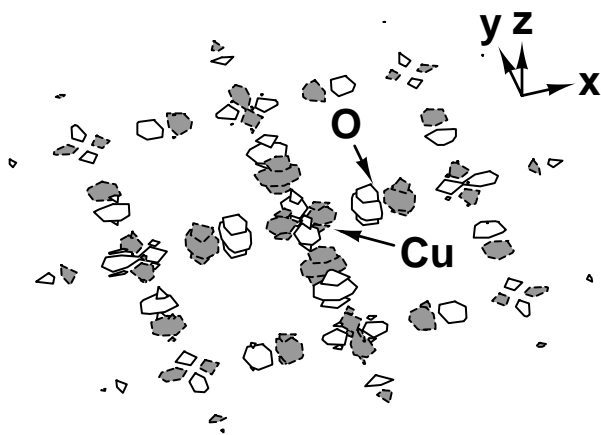}}
\caption {
Highest occupied molecular orbital for the
Cu$_9$O$_{42}$/Cu$_{12}$La$_{50}$ cluster.
\label{homocuo}}}
\end{figure}

\begin{figure}
{\centerline{\includegraphics[height=6cm]{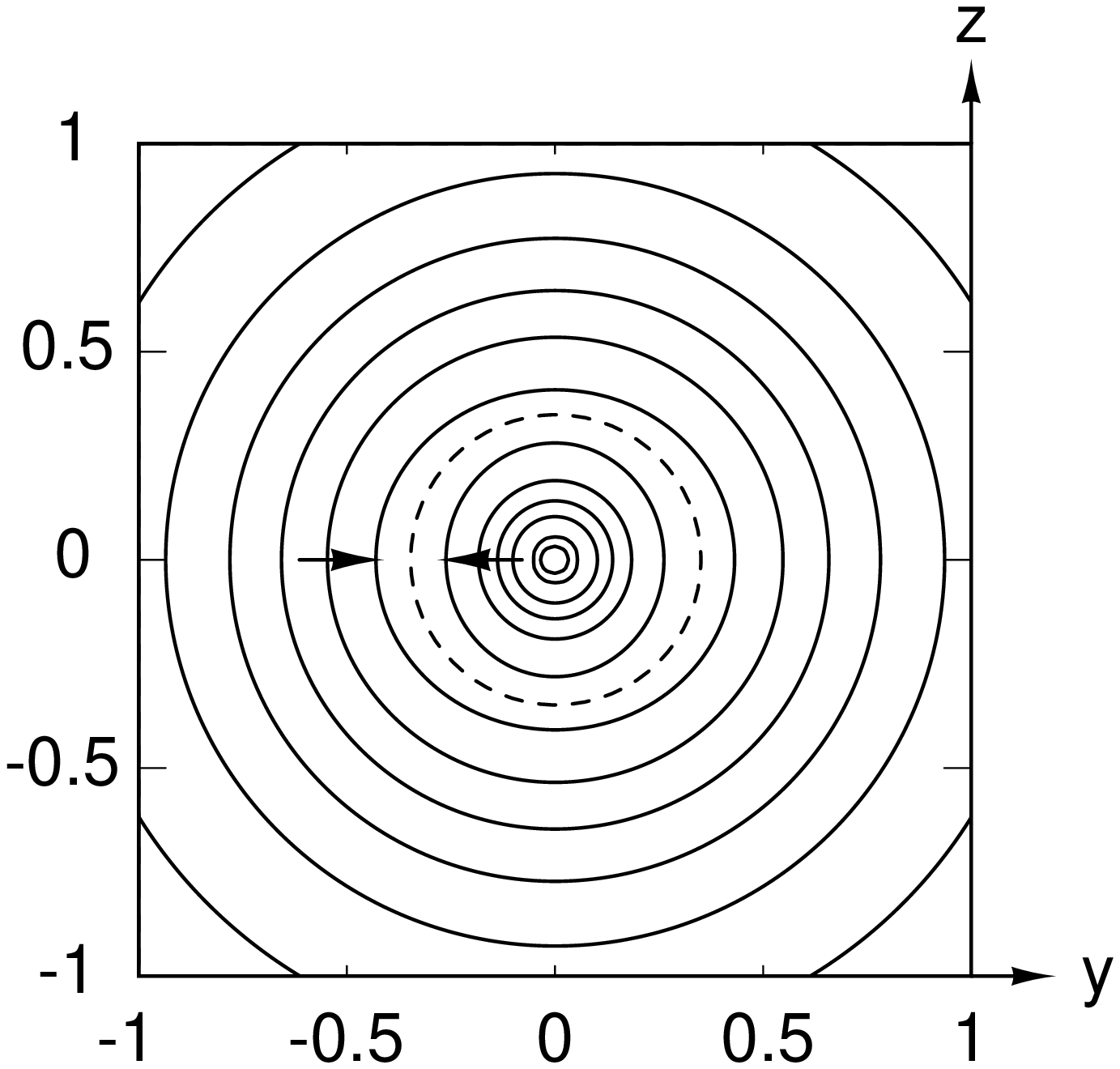}}
\caption {
Density distribution of the $2p$ oxygen electrons in a $yz$-plane
perpendicular to the Cu-O-Cu connection line and through the oxygen atom.
The arrows point along increasing densities. The equidensity
lines close the density maxima show that these maxima are larger in the $y$
than in the $z$ direction.
\label{denasyo}}}
\end{figure}

\begin{thebibliography}{99}  
\bibitem{adrian} F. J. Adrian, Phys. Rev. B {\bf 38}, 2426 (1988).
\bibitem{Shimizu}T. Shimizu, H. Yasuoka, T. Imai, T. Tsuda, T. Takabatake, Y.
  Nakazawa, and M. Ishikawa, J. Phys. Soc. Jpn. {\bf 57}, 2494 (1988).
\bibitem{pennington}C. H. Pennington, D. J. Durand, C. P. Slichter, J. P. 
        Rice, E. D. Bukowski, and D. M. Ginsberg, Phys. Rev. B {\bf 39}, 
        2902 (1989).
\bibitem{Garcia}M. E. Garcia and K. H. Bennemann, Phys. Rev. B {\bf 40},
  8809 (1989).
\bibitem{kat} K. M\"uller, M. Mali, J. Roos and D. Brinkmann, 
        Physica C {\bf 162-164}, 173 (1989).
\bibitem{takigawa} M. Takigawa, P. C. Hammel, R. H. Heffner, Z. Fisk, 
        K. C. Ott, and J. D. Thompson, Phys. Rev. Lett., {\bf 63}, 1865 (1989).
\bibitem{hanzawa} K. Hanzawa, F. Komatsu and K. Yosida, J. Phys. Soc. Jpn.,
              {\bf 59}, 3345 (1990).
\bibitem {Stern} R. M. Sternheimer, Phys. Rev. {\bf 95}, 736 (1954).
\bibitem {Stern2} M. H. Cohen and F. Reif, Solid State Physics, Vol. 5,
  ed. by F. Seitz and D. Turnbull, Academic N. Y. (1957).
\bibitem{ohta} Y. Ohta, W. Koshibae and S. Maekawa, J. Phys. Soc. Jpn.,
              {\bf 61}, 2198 (1992).
\bibitem{zheng} G. Zheng, Y. Kitaoka, K. Ishida and K. Asayama, J. Phys. Soc.
                Jpn., {\bf 64}, 2524 (1995).
\bibitem{asayama} K. Asayama, Y. Kitaoka, G.-q. Zheng, K. Ishida, and
            K. Magishi, Physica B {\bf 223} \& {\bf 224}, 478 (1996).
\bibitem{Kupcic} I. Kup$\check{\rm c}$i\'c, S. Bari$\check{\rm s}$i\'c, 
             and E. Tuti\'s, Phys. Rev. B {\bf 57}, 8590 (1998).
\bibitem{walstedt} R. E. Walstedt and S-W. Cheong, 
                    Phys. Rev. B {\bf 64}, 014404 (2001).
\bibitem {pc}P. C. Schmidt, private communication.
\bibitem{La}P. H\"usser, H. U. Suter, E. P. Stoll, and P. F. Meier, 
            Phys. Rev. B {\bf 61}, 1567 (2000).
\bibitem{Y}S. Renold, S. Pliber\v{s}ek, E. P. Stoll, T. A. Claxton, and
  P. F. Meier, Eur. Phys. J. B, {\bf23}, 3 (2001). 
\bibitem{pw1}K. Schwarz, C. Ambrosch-Draxl, and P. Blaha, 
  Phys. Rev. B {\bf 42}, 2051 (1990).
\bibitem{pw2}J. Yu, A. J. Freeman, R. Podloucky, P. Herzig, and P. Weinberger,
  Phys. Rev. B {\bf 43}, 532 (1991).
\bibitem{lattice}   {\em Copper Oxide Superconductors},
 Ch. P. Poole, T. Datta and H.A. Farach, (Wiley-Intescience, New York, 1988).
\bibitem{mulliken} R. S. Mulliken, J. Chem. Phys. {\bf 23}, 1833 (1955).
\bibitem{politzer} E. W. Stout, Jr. and P. Politzer, 
           Theoret. Chim. Acta {\bf 12}, 379 (1968).
\bibitem{lowdin} P.-O. L\"owdin, J. Chem. Phys. {\bf 18}, 365 (1950).
\bibitem {abragam} A. Abragam, {\it The Principles of Nuclear Magnetism}
  (Oxford University Press, New York, 1961).
\bibitem{Leipzig}P. F. Meier, T. A. Claxton, P. H\"usser, S. Pliber\v{s}ek,
            and E. P. Stoll, Z. Naturforsch. {\bf 55 a}, 247 (2000).
\end{thebibliography}
\end{document}